\documentclass[aps,english,preprint,nofootinbib]{revtex4}
\pdfoutput=1

\usepackage{graphicx}
\usepackage{hyperref}
\usepackage{amsmath, amssymb}
\usepackage{babel}
\usepackage{color}
\usepackage{slashed}   
\usepackage[utf8]{inputenc}
\usepackage{multirow}

\begin{document}

\title{Constraints on Disconnected Contributions in $\pi\pi$ Scattering}

\author{N. Ripunjay Acharya$^{1}$, Feng-Kun Guo$^{2,3}$, Ulf-G. Mei\ss{}ner$^{1,4,5}$ and Chien-Yeah Seng$^{1}$}

\affiliation{$^1$Helmholtz-Institut f\"ur Strahlen- und Kernphysik and Bethe Center for Theoretical Physics,\\ Universit\"at Bonn, 53115 Bonn, Germany \\
	$^2$CAS Key Laboratory of Theoretical Physics,
	Institute of Theoretical Physics,\\ Chinese Academy of Sciences,
	Beijing 100190, China\\
	$^3$School of Physical Sciences,
	University of Chinese Academy of Sciences,
	Beijing 100049, China\\
	$^4$Institute for Advanced Simulation, Institut f\"ur Kernphysik and J\"ulich Center for Hadron Physics, Forschungszentrum J\"ulich, 52425 J\"ulich, Germany\\
	$^5$Tbilisi State  University,  0186 Tbilisi, Georgia }

\date{\today}

\begin{abstract}

The accuracy of the lattice QCD computation of hadron-hadron scattering at low isospin depends critically on the ability to compute correlation functions with fermionic disconnected Wick contractions. This happens, for instance, in  isospin $I=0$ $\pi\pi$ scattering, which receives contributions from rectangular and vacuum types of contractions among other easier calculable ones.  Combining L\"{u}scher's formula and partially-quenched chiral perturbation theory, we provide  precise theory predictions of the discrete energy levels extracted from specific linear combinations of lattice correlation functions corresponding to various types of contractions. Expressions are provided for extracting the unphysical low-energy constants in the partially-quenched chiral perturbation theory from the energy levels for these contractions. The predictions for the rectangular and vacuum contractions may serve as solid tests of the accuracy for existing and future lattice studies of $\pi\pi$ scattering.

\end{abstract}

\maketitle

\section{Introduction}

Elastic $\pi\pi$ scattering is among the simplest processes governed by Quantum Chromodynamics (QCD) in its nonperturbative regime, yet it contains a rich amount of interesting physics ranging from  spontaneous chiral symmetry breaking to low-lying meson resonances. First-principle studies of the scattering process thus play a critical role in refining our understanding of the properties of strong interactions at low energies. In particular, there has been a continuous effort since the 1990s in the simulation of the $\pi\pi$ scattering using lattice QCD that covers all isospin channels~\cite{Sharpe:1992pp,Kuramashi:1993ka,Gupta:1993rn,Fukugita:1994ve,
	Aoki:2002in,Du:2004ib,Chen:2005ab,Li:2007ey,Beane:2007xs,Aoki:2007rd,Feng:2009ij,Feng:2010es,Dudek:2010ew,Lang:2011mn,Aoki:2011yj,Beane:2011sc,Dudek:2012gj,Pelissier:2012pi, 
	Dudek:2012xn,Fu:2013ffa,Feng:2014gba,Bai:2015nea,Helmes:2015gla,Wilson:2015dqa,Bali:2015gji,Bulava:2016mks,Guo:2016zos,Briceno:2016mjc,
	Liu:2016cba,Alexandrou:2017mpi}.

Studies of  $\pi\pi$ scattering in lattice QCD involve the computation of four-point correlation functions. Quantized energy levels are fixed from the time-dependence of these correlation functions, and the scattering observables are extracted from these volume-dependent energies using
L\"uscher's method~\cite{Luscher:1986pf,Luscher:1990ux}. Pion-pion scattering amplitudes for total isospin $I=1$ and $I=2$  are relatively easy to compute.
However, it is well-known that the lattice computation of $I=0$ $\pi\pi$ scattering is significantly more challenging due to the existence of quark contraction diagrams in which the quark propagators begin and end at the same temporal point on the lattice: These diagrams are known as ``disconnected diagrams" for historical reasons. In particular, there are the so-called vacuum-type contractions, in which the quark lines in the source and those in the sink are separated from each other. Such contractions possess a low signal-to-noise ratio, and usually require special algorithms (such as the stochastic LapH quark smearing method~\cite{Peardon:2009gh,Morningstar:2011ka} that provides all-to-all propagators) that are computationally demanding to achieve good statistical precision. For instance, in Ref.~\cite{Liu:2016cba} it is clear that the disconnected correlation functions suffer from larger noise than the connected ones, while Ref.~\cite{Briceno:2016mjc} achieved better precision using a larger basis of interpolating fields but their computation is significantly more expensive.

In the meantime, corresponding theoretical tools have been developed in an attempt to model the properties of disconnected diagrams and eventually to serve as a gauge for any possible future improvement of relevant lattice computational techniques. They generally involve extensions of the flavor sector in QCD such that the contraction diagrams of interest can be singled out by choosing appropriately the quark flavors in the external scattering states. The first attempt of such a kind was taken in Ref.~\cite{Guo:2013nja} that made use of an SU(4) chiral perturbation theory (ChPT) to describe the separation of different contraction diagrams at leading order (LO) in the chiral expansion. Follow-up works were carried out in Ref.~\cite{Acharya:2017zje} based on the SU$(4|2)$ partially-quenched chiral perturbation theory (PQChPT)~\cite{Bernard:1993sv,Sharpe:1999kj,Sharpe:2000bc,Sharpe:2001fh,Bernard:2013kwa,Sharpe:2006pu,Golterman:2009kw} that fully preserves the internal dynamics of the ordinary two-flavor ChPT. Scattering amplitudes corresponding to each individual contraction were computed to the next-to-leading order (NLO), and analytic expressions for static quantities such as scattering lengths were derived. The problem with these calculations, however, is that the results were obtained in a field theory in the infinite volume, whereas what one obtains on the lattice are always discrete energy levels because of the  finite volumes. It is therefore unclear how the energy levels extracted from the lattice correlation functions of definite contractions are related to the quantities (such as scattering lengths) derived in Ref.~\cite{Acharya:2017zje}.

\begin{figure}[tb]
	\begin{centering}
		\includegraphics[scale=0.2]{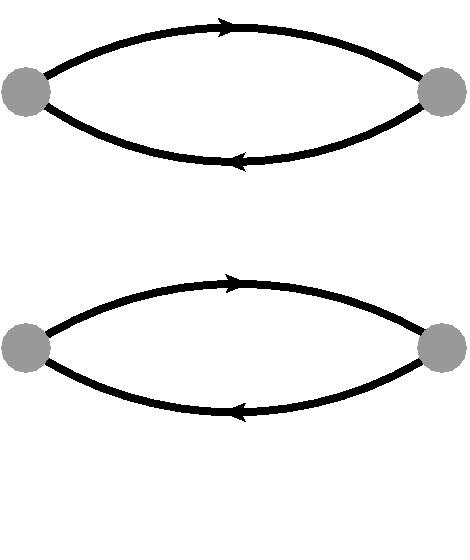}\hfill
		\includegraphics[scale=0.2]{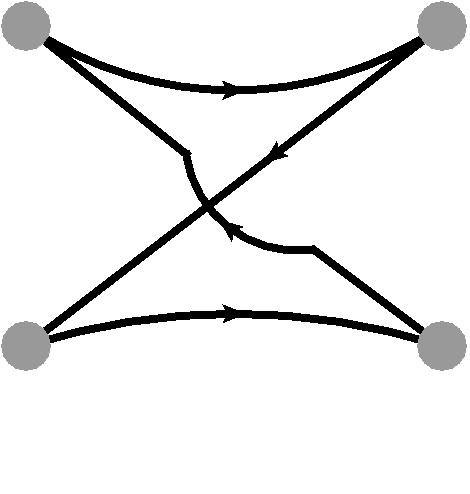}\hfill
		\includegraphics[scale=0.2]{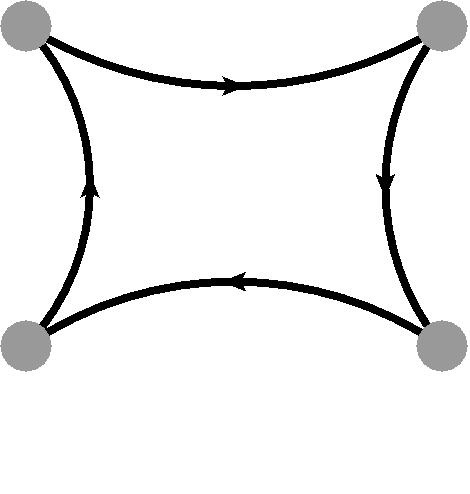}\hfill
		\includegraphics[scale=0.2]{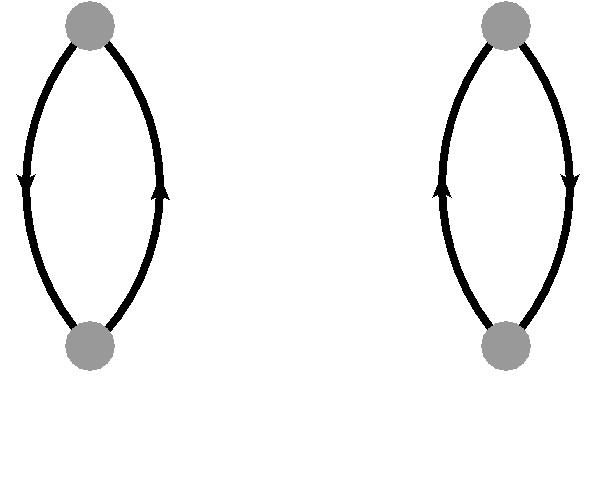}
		\par\end{centering}
	\caption{\label{fig:contraction}(From left to right) direct (D), crossed (C), rectangular (R) and vacuum (V) contractions in a generic $\pi\pi$ scattering amplitude.}
\end{figure}
We provide in this paper a satisfactory solution to the above-mentioned problem by combining PQChPT with the single-channel L\"{u}scher's formula~\cite{Luscher:1986pf,Luscher:1990ux}. First, we introduce auxiliary valence flavors, which are degenerate with the sea up and down quarks (we consider the isospin symmetric case), using the graded group SU$(4|2)$. This enables us to have enough channels to separate various types of contractions, which include direct (D), crossed (C), rectangular (R) and vacuum (V) types, see Fig.~\ref{fig:contraction}. Because of the degeneracy of all the sea and valence quarks in the setting of our theory, by diagonalizing a matrix for multi-channel $S$-wave scattering amplitudes in SU$(4|2)$, we obtain four linear combinations of scattering amplitudes that are effectively single-channel. Two of them consist of only the connected contractions (D and C) and the other two involve the R and V contractions as well. The single-channel L\"{u}scher's formula then provides the relation between the discrete energy level of each combination in a finite volume and the corresponding low-energy scattering observables, such as the scattering length and effective range, in the infinite volume.  The latter are also calculated in PQChPT up to NLO and are expressed in terms of the low-energy constants (LECs) of the theory. Since the discrete energies involving only the connected contractions are easily obtainable on the lattice, they can be used to fix the poorly-known ``unphysical" LECs in the theory, which have so far only been fitted to the meson masses and decay constants in the lattice calculations of Ref.~\cite{Boyle:2015exm}. After doing so, the two remaining energy levels involving disconnected diagrams can be predicted in PQChPT. The comparison between the energy levels extracted from lattice correlation functions and the theory predictions then serves as a powerful gauge for the accuracy of the former, because the latter are model-independent results that must be obeyed (up to corrections from higher orders in the chiral power counting and from finite lattice spacing) by any lattice setup that satisfies unitarity.

This paper is organized as follows. In Section~\ref{sec:Luescher}, we review the single-channel L\"{u}scher's formula that relates the energy levels in a finite volume to the scattering observables in the infinite volume. In Section~\ref{sec:effective}, we derive four linear combinations of the $S$-wave $\pi\pi$ amplitudes that comprise effective single-channel scattering amplitudes in SU$(4|2)$. The PQChPT results of their respective scattering length and effective range are summarized in Section~\ref{sec:ere}. In Section~\ref{sec:numerics}, we make use of existing lattice data on connected diagrams in $\pi\pi$ scattering to fit a subset of the so-called ``unphysical" LECs in SU$(4|2)$, and thus give our predictions of the two remaining energy levels that depend on the disconnected diagrams. We sum up with conclusions in Section~\ref{sec:conclusion}. 

\section{Single-Channel L\"{u}scher's Formula}
\label{sec:Luescher}

We start by reviewing  L\"{u}scher's formula for single-channel partial-wave amplitudes. A single-channel scattering amplitude $T(s,t,u)$ between two non-identical particles in the center of mass (CM) frame can be expanded in terms of its partial-wave amplitudes $\{T_l(E)\}$,
\begin{equation}
T(s,t,u)=\sum_{l=0}^{\infty}(2l+1)T_l(E)P_l(\cos\theta), \label{eq:PWE}
\end{equation}
where the $P_l(\cos\theta)$ are Legendre polynomials, $E=\sqrt{s}$ is the CM energy, and $\theta$ is the scattering angle. The partial-wave amplitudes satisfy the following unitarity relation:
\begin{equation}
\mathrm{Im}T_l(E)=\Theta(E-m_1-m_2)\frac{p}{8\pi E}|T_l(E)|^2,\label{eq:unitarity}
\end{equation}
with $p$ the CM momentum. This allows us to parameterize the partial-wave amplitudes in terms of their phase shift $\delta_l$ as
\begin{equation}
T_l(E)=\frac{8\pi E}{p\cot\delta_l(E)-ip}.\label{eq:phaseshift}
\end{equation}
From now on we shall restrict ourselves to the special case where $m_1=m_2=M_\pi$, and thus $p=\sqrt{E^2-4M_\pi^2}/2$.

Lattice studies of scattering processes are performed in finite spatial volumes characterized by the lattice size $L$. Four-point correlation functions are computed, and the measured outcomes are discrete energy levels which correspond to the poles of the finite-volume $T$-matrix. 
L\"{u}scher \cite{Luscher:1986pf,Luscher:1990ux} derived a relation between the volume dependence of two-particle energies in the finite volume and the partial-wave amplitudes in the infinite volume. We consider here the $S$-wave case. The relation is recast into an equivalent but more understandable form in Ref.~\cite{Doring:2011vk} which we shall adopt here\footnote{Notice that the right-hand side of our Eq. \eqref{eq:Luscher} has a sign difference with that in Ref.~\cite{Doring:2011vk} because our partial-wave amplitudes are defined to have a positive imaginary part above threshold.},
\begin{equation}
T_0^{-1}(E_0)=-\Delta G(E_0),\label{eq:Luscher}
\end{equation}
which holds for the eigenenergies of the two-particle system in a finite volume, denoted by $E_0$. Here $\Delta G(E)=\tilde{G}(E)-G(E)$, where
\begin{equation}
G(E)=\int_{|\vec{q}|<\Lambda}\frac{d^3q}{(2\pi)^3\omega_\pi}\frac{1}{E^2-4\omega_\pi^2+i\varepsilon},\qquad 
\tilde{G}(E)=\frac{1}{L^3}\sum_{|\vec{q}|<\Lambda}\frac{1}{\omega_\pi}\frac{1}{E^2-4\omega_\pi^2}
\end{equation}
with $\omega_\pi=\sqrt{M_\pi^2+\vec{q}^2}$, are the infinite-volume and finite-volume versions of the $s$-channel 2-point scalar loop function, both regularized at the ultraviolet (UV) end by a hard cutoff $\Lambda$. Notice that the discrete loop momentum $\vec{q}$ in $\tilde{G}(E)$ is given by $\vec{q}=(2\pi/L)\vec{n}$ with $\vec{n}\in \mathbb{Z}^3$. Also, the function $\Delta G(E)$ is UV finite and insensitive to the cutoff $\Lambda$ as long as $\Lambda\gg M_\pi$, as the UV part gets cancelled between $G$ and $\tilde G$. 

When $p$ is small, one could perform the following effective-range expansion to $p\cot\delta_0$,\footnote{The scattering length defined here has a sign difference to that in Ref.~\cite{Helmes:2015gla}.}
\begin{equation}
p\cot\delta_0=-\frac{1}{a_0}+\frac{1}{2}r_0p^2+\mathcal{O}(p^4).\label{eq:EFexpand}
\end{equation}
The right-hand side is valid both above and below threshold, for the latter case we simply have $p^2<0$. The quantities $a_0$ and $r_0$ are known as the ($S$-wave) scattering length and effective range, respectively. Substituting Eqs.~\eqref{eq:phaseshift} and \eqref{eq:EFexpand} into Eq.~\eqref{eq:Luscher} yields
\begin{equation}
-\frac{1}{a_0}+\frac{1}{2}r_0p^2+...=-8\pi E_0\Delta G(E_0)+ip\,,\label{eq:central}
\end{equation}
which is the starting point of our finite-volume analysis. Notice that the right side is always real: Above threshold, the imaginary part of $-8\pi E_0\Delta G(E_0)$ cancels with $ip$, whereas below threshold one performs the analytic continuation $p=i\sqrt{4M_\pi^2-E_0^2}/2$ \cite{Doring:2011vk}. This equation serves two purposes: (1) If $E_0$ could be extracted from lattice for several values of $L$, then one could use them to fit the scattering length and effective range $\{a_0,r_0\}$ to determine the unknown theoretical parameters (LECs) within these quantities, and (2) if the values of $\{a_0,r_0\}$ could be calculated from theory, then one could use Eq.~\eqref{eq:central} to predict the discrete energy $E_0$, which can be used to test the accuracy of the corresponding lattice calculation.

\section{Effective Single-Channel $S$-wave Amplitudes}
\label{sec:effective}

The $I=0$ $\pi\pi$-scattering amplitude consists of four types of contraction diagrams as depicted in Fig.~\ref{fig:contraction}: D, C, R and V as well as their $t\leftrightarrow u$ crossings. As described in the Introduction, D and C are connected diagrams that could be obtained rather easily on the lattice, whereas R needs more efforts and V is the most challenging. Each contraction can be rigorously defined in terms of ``physical" scattering amplitudes between pseudo-Nambu--Goldstone bosons (pNGBs) in a degenerate SU$(4|2)$ partially-quenched QCD (PQQCD), for which we refer the readers to Ref.~\cite{Acharya:2017zje} for details,
\begin{eqnarray}
T^D(s,t,u)&\equiv &T_{(u\bar{d})(j\bar{k})\rightarrow (u\bar{d})(j\bar{k})}(s,t,u) ,\nonumber\\
T^C(s,t,u)&\equiv &T_{(u\bar{d})(j\bar{k})\rightarrow (u\bar{k})(j\bar{d})}(s,t,u) ,\nonumber\\
T^R(s,t,u)&\equiv &T_{(j\bar{u})(u\bar{k})\rightarrow (j\bar{d})(d\bar{k})}(s,t,u) ,\nonumber\\
T^V(s,t,u)&\equiv &T_{(u\bar{d})(d\bar{u})\rightarrow (j\bar{k})(k\bar{j})}(s,t,u),
\end{eqnarray}
where $u,d,j$ and $k$ refer to the quark flavors in the SU$(4|2)$ theory, and $(u\bar d)$ and so on refer to the corresponding pNGBs.

None of the amplitudes above describes a single-channel scattering in SU$(4|2)$. However, it is straightforward to construct linear combinations of amplitudes that are effectively single-channel. To demonstrate the mechanism, let us consider a multi-channel scattering process that involves only the D, C and R contractions. This can be done by including only five channels as follows: $(j\bar{u})(u\bar{k})$, $(j\bar{d})(d\bar{k})$, $\pi^0(j\bar{k})$, $\tilde{\eta}(j\bar{k})$ and $\tilde{\phi}(j\bar{k})$, where
\begin{equation}
\pi^0=\frac{1}{\sqrt{2}}\left(u\bar{u}-d\bar{d}\right),\:\:\:
\tilde{\eta}=\frac{1}{\sqrt{6}}\left(u\bar{u}+d\bar{d}-2j\bar{j}\right),\:\:\:
\tilde{\phi}=\frac{1}{\sqrt{12}}\left(u\bar{u}+d\bar{d}+j\bar{j}-3k\bar{k}\right)
\end{equation}
in the flavor space. The $S$-wave amplitudes in this coupled-channel system are then a $5\times 5$ symmetric matrix, 
\begin{equation}
T_{0}=\left(\begin{array}{ccccc}
T_{0}^{D}+T_{0}^{R} & T_{0}^{R} & \frac{1}{\sqrt{2}}T_{0}^{C} & \frac{1}{\sqrt{6}}\left(T_{0}^{C}-2T_{0}^{R}\right) & \frac{1}{2\sqrt{3}}\left(T_{0}^{C}-2T_{0}^{R}\right)\\
T_{0}^{R} & T_{0}^{D}+T_{0}^{R} & -\frac{1}{\sqrt{2}}T_{0}^{C} & \frac{1}{\sqrt{6}}\left(T_{0}^{C}-2T_{0}^{R}\right) & \frac{1}{2\sqrt{3}}\left(T_{0}^{C}-2T_{0}^{R}\right)\\
\frac{1}{\sqrt{2}}T_{0}^{C} & -\frac{1}{\sqrt{2}}T_{0}^{C} & T_{0}^{D} & 0 & 0\\
\frac{1}{\sqrt{6}}\left(T_{0}^{C}-2T_{0}^{R}\right) & \frac{1}{\sqrt{6}}\left(T_{0}^{C}-2T_{0}^{R}\right) & 0 & T_{0}^{D}+\frac{2}{3}\left(T_{0}^{C}+T_{0}^{R}\right) & \frac{1}{3\sqrt{2}}\left(2T_{0}^{R}-T_{0}^{C}\right)\\
\frac{1}{2\sqrt{3}}\left(T_{0}^{C}-2T_{0}^{R}\right) & \frac{1}{2\sqrt{3}}\left(T_{0}^{C}-2T_{0}^{R}\right) & 0 & \frac{1}{3\sqrt{2}}\left(2T_{0}^{R}-T_{0}^{C}\right) & T_{0}^{D}+\frac{5}{6}T_{0}^{C}+\frac{1}{3}T_{0}^{R}
\end{array}\right),
\end{equation}
which can be diagonalized through an orthogonal transformation to give
\begin{equation}
T'_0=UT_0U^T=\mathrm{diag}\left(T_0^D+T_0^C,T_0^D+T_0^C,T_0^D+T_0^C,T_0^D-T_0^C,T_0^D-\frac{1}{2}T_0^C+3T_0^R\right).
\end{equation}
Furthermore, since all pNGBs are degenerate, each of the diagonal entries of $T'_0$ (among which only three of them are independent) is by itself an effective single-channel $S$-wave amplitude that satisfies the unitarity relation in Eq.~\eqref{eq:unitarity}, and can be parameterized according to
Eq.~\eqref{eq:phaseshift}.

The treatment above can be generalized to include the vacuum contraction, but then it will involve the diagonalization of a $12\times 12$ matrix. In fact, such a generalization is not necessary because we already know the fourth linear combination. It is just the $I=0$ $S$-wave amplitude,\footnote{While the lattice community usually adopts the opposite sign convention for the crossed and rectangular contractions: $T_0^{I=0}=2T_0^D+T_0^C-6T_0^R+3T_0^V$, our sign convention is more natural in terms of the interpretation of different contractions as physical scattering amplitudes in SU$(4|2)$.}
\begin{equation}
T_0^{I=0}=2T_0^D-T_0^C+6T_0^R+3T_0^V.
\end{equation} 
The only complication is that we need to rescale it by a factor 1/2 to match the normalization in Eq. \eqref{eq:unitarity}, due to the fact that $T_0^{I=0}$ involves identical particles. Up to this point, we have successfully constructed four linear combinations of D, C, R and V that give rise to effective single-channel $S$-wave amplitudes in SU$(4|2)$:
\begin{eqnarray}
T_0^\alpha&=&T_0^D+T_0^C,\nonumber\\
T_0^\beta&=&T_0^D-T_0^C,\nonumber\\
T_0^\gamma&=&T_0^D-\frac{1}{2}T_0^C+3T_0^R,\nonumber\\
T_0^\delta&=&T_0^D-\frac{1}{2}T_0^C+3T_0^R+\frac{3}{2}T_0^V,
\end{eqnarray}
among which $T_0^\alpha$ and $T_0^\delta$ correspond (up to a constant factor of 2) to the $I=2$ and $I=0$ $\pi\pi$ scattering amplitude, respectively. The implications of this construction on lattice are as follows. Suppose $\tilde{C}^{I=0}(\tau)=2\tilde{C}^D(\tau)-\tilde{C}^C(\tau)+6\tilde{C}^R(\tau)+3\tilde{C}^V(\tau)$ is the lattice correlation function used to extract the discrete energy levels in the $I=0$ $\pi\pi$ scattering, then the following linear combinations of $\tilde{C}^X(\tau)$ ($X=D,C,R,V$) will decay as a single exponential function at large Euclidean time $\tau$:
\begin{eqnarray}
\tilde{C}^\alpha(\tau)\equiv 2\tilde{C}^D(\tau)+2\tilde{C}^C(\tau)&\sim &\exp\{-E_0^\alpha\tau\},\nonumber\\
\tilde{C}^\beta(\tau)\equiv 2\tilde{C}^D(\tau)-2\tilde{C}^C(\tau)&\sim &\exp\{-E_0^\beta\tau\},\nonumber\\
\tilde{C}^\gamma(\tau)\equiv 2\tilde{C}^D(\tau)-\tilde{C}^C(\tau)+6\tilde{C}^R(\tau)&\sim &\exp\{-E_0^\gamma\tau\},\nonumber\\
\tilde{C}^\delta(\tau)\equiv 2\tilde{C}^D(\tau)-\tilde{C}^C(\tau)+6\tilde{C}^R(\tau)+3\tilde{C}^V(\tau)&\sim & \exp\{-E_0^\delta\tau\},
\label{eq:Edef}
\end{eqnarray}
where the energy levels $E^i_0$ ($i=\alpha,\beta,\gamma,\delta$) correspond to the poles of the finite-volume $T_0^i(E)$, and thus satisfy the single-channel L\"{u}scher formula in Eq.~\eqref{eq:Luscher}. We wish also to point out the group theoretical origin of the four linear combinations. They are nothing but elements of the four irreducible representations resulting from the symmetric product of two 15-plets (that represent the two pNGBs) in SU(4)~\cite{Sharpe:1992pp},
\begin{equation}
(\textbf{15}\otimes\textbf{15})_\mathrm{symm}=\textbf{1}\oplus\textbf{15}\oplus\textbf{20}\oplus\textbf{84}~.
\end{equation}

\section{PQChPT Predictions of Scattering Lengths and Effective Ranges}
\label{sec:ere}

The expansion of $\mathrm{Re}\left[\left(T_0^i(E)\right)^{-1}\right]$ around the two-pion threshold gives the scattering length $a_0^i$ and the effective range $r_0^i$ for $i=\alpha,\beta,\gamma,\delta$. The PQChPT predictions of these quantities up to NLO can be obtained through simple manipulations of the results in Ref.~\cite{Acharya:2017zje}. The outcomes are \footnote{We shall take this opportunity to point out that the analytic expression for the $I=0$ effective range given in Appendix A of Ref.~\cite{Liu:2016cba} is incorrect.}
\begin{eqnarray}
a_0^\alpha&=&\frac{M_\pi}{16\pi F_\pi^2}\left[1+\frac{M_\pi^2}{\pi^2F_\pi^2}\left(-\frac{\bar{l}_1}{12}-\frac{\bar{l}_2}{6}+\frac{\bar{l}_3}{32}+\frac{\bar{l}_4}{8}-\frac{1}{32}\right)\right],\nonumber\\
r_0^\alpha&=&\frac{48\pi F_\pi^2}{M_\pi^3}\left[1+\frac{M_\pi^2}{\pi^2F_\pi^2}\left(\frac{\bar{l}_1}{12}+\frac{\bar{l}_2}{18}-\frac{7\bar{l}_3}{96}-\frac{\bar{l}_4}{8}+\frac{31}{288}\right)\right],\label{eq:aralpha}
\end{eqnarray} 
\begin{eqnarray}
a_0^\beta&=&-\frac{M_\pi}{16\pi F_\pi^2}\left[1-4\mu_\pi+\frac{M_\pi^2}{F_\pi^2}\left(\frac{\bar{l}_1}{12\pi^2}+\frac{\bar{l}_2}{6\pi^2}-\frac{\bar{l}_3}{32\pi^2}-\frac{\bar{l}_4}{8\pi^2}-96L_0^\mathrm{PQ,r}\right.\right.\nonumber\\
&&\left.\left.-32L_3^\mathrm{PQ,r}+32L_5^\mathrm{PQ,r}-32L_8^\mathrm{PQ,r}-\frac{3}{32\pi^2}\right)\right],\nonumber\\
r_0^\beta&=&-\frac{48\pi F_\pi^2}{M_\pi^3}\left[1+\frac{100}{9}\mu_\pi+\frac{M_\pi^2}{F_\pi^2}\left(-\frac{\bar{l}_1}{12\pi^2}-\frac{\bar{l}_2}{18\pi^2}+\frac{7\bar{l}_3}{96\pi^2}+\frac{\bar{l}_4}{8\pi^2}+\frac{160}{3}L_0^\mathrm{PQ,r}\right.\right.\nonumber\\
&&\left.\left.+32L_3^\mathrm{PQ,r}-\frac{160}{3}L_5^\mathrm{PQ,r}+\frac{224}{3}L_8^\mathrm{PQ,r}+\frac{61}{288\pi^2}\right)\right],
\end{eqnarray}
\begin{eqnarray}
a_0^\gamma&=&-\frac{7M_\pi}{32\pi F_\pi^2}\left[1-\frac{18}{7}\mu_\pi+\frac{M_\pi^2}{F_\pi^2}\left(\frac{\bar{l}_1}{42\pi^2}+\frac{\bar{l}_2}{21\pi^2}-\frac{\bar{l}_3}{112\pi^2}-\frac{\bar{l}_4}{28\pi^2}+\frac{144}{7}L_0^\mathrm{PQ,r}\right.\right.\nonumber\\
&&\left.\left.+\frac{48}{7}L_3^\mathrm{PQ,r}+\frac{48}{7}L_5^\mathrm{PQ,r}+\frac{48}{7}L_8^\mathrm{PQ,r}+\frac{5}{56\pi^2}\right)\right],\nonumber\\
r_0^\gamma&=&-\frac{288\pi F_\pi^2}{49M_\pi^3}\left[1-\frac{110}{63}\mu_\pi+\frac{M_\pi^2}{F_\pi^2}\left(\frac{\bar{l}_1}{126\pi^2}+\frac{17\bar{l}_2}{189\pi^2}+\frac{25\bar{l}_3}{1008\pi^2}-\frac{\bar{l}_4}{84\pi^2}\right.\right.\nonumber\\
&&\left.\left.+\frac{592}{21}L_0^\mathrm{PQ,r}+\frac{496}{21}L_3^\mathrm{PQ,r}+\frac{16}{7}L_5^\mathrm{PQ,r}-\frac{400}{21}L_8^\mathrm{PQ,r}-\frac{25}{189\pi^2}\right)\right],\label{eq:argamma}
\end{eqnarray}
\begin{eqnarray}
a_0^\delta&=&-\frac{7M_\pi}{32\pi F_\pi^2}\left[1+\frac{M_\pi^2}{\pi^2F_\pi^2}\left(\frac{5\bar{l}_1}{84}+\frac{5\bar{l}_2}{42}-\frac{5\bar{l}_3}{224}+\frac{\bar{l}_4}{8}+\frac{5}{32}\right)\right],\nonumber\\
r_0^\delta&=&-\frac{288\pi F_\pi^2}{49M_\pi^3}\left[1+\frac{M_\pi^2}{\pi^2F_\pi^2}\left(\frac{11\bar{l}_1}{84}+\frac{43\bar{l}_2}{378}+\frac{125\bar{l}_3}{2016}-\frac{\bar{l}_4}{8}-\frac{479}{864}\right)\right],
\label{eq:ardelta}\end{eqnarray}
where $\mu_\pi=\left(M_\pi^2/(32\pi^2 F_\pi^2)\right)\ln(M_\pi^2/\mu^2)$, with $\mu$ the renormalization scale and $M_\pi,F_\pi$ the scale-independent physical pion mass and decay constant ($F_\pi^{\mathrm{phy}} = 92.1$ MeV). The constants $\{\bar{l}_1\}$ and $\{L_i^\mathrm{PQ,r}\}$ are known as the scale-independent physical and scale-dependent unphysical LECs in SU$(4|2)$ PQChPT, respectively. 
The former ones enter  processes in the real world, but their values are $M_\pi$-dependent: $\bar{l}_i(M_\pi)=\bar{l}_i^\mathrm{phy}-\ln(M_\pi^2/M_{\pi,\mathrm{phy}}^2)$ where $M_{\pi,\mathrm{phy}}\approx 138$ MeV is the physical pion mass. The latter ones are $M_\pi$-independent, but they can only show up either when unphysical quark flavors appear in the external states, or when the valence and sea quark masses in an ordinary QCD are intentionally made different. In the second case, $\{L_{5,8}^\mathrm{PQ,r}\}$ affect the pion mass and decay constant at NLO, which allows for a fairly precise determination of their values on the lattice as shown in Ref.~\cite{Boyle:2015exm}. On the other hand, the effect of $\{L_{0,3}^\mathrm{PQ,r}\}$ on the pion mass and decay constant appears only at NNLO so that their values are poorly constrained. Finally, we note that the $\mu$-dependence in $\mu_\pi$ and $L_i^\mathrm{PQ,r}$ always cancel each other to make the final expressions of $\{a_0^i,r_0^i\}$ scale-independent.

Until now we have presented all theoretical ingredients needed in this work.
The usefulness of the results above is as follows. First, the energy levels $E_0^{\alpha,\beta}$ that could be obtained rather precisely from the D and C correlation functions on lattice can be used to determine a number of physical and unphysical LECs in PQChPT through the fitting of $\{a_0^{\alpha,\beta},r_0^{\alpha,\beta}\}$. In particular, $a_0^\beta$ and $r_0^\beta$ depend on different linear combinations of the two poorly-constrained unphysical LECs $L_0^\mathrm{PQ,r}$ and $L_3^\mathrm{PQ,r}$, so that the lattice calculation of $E_0^\beta$ with several values of $L$ will help pinning down these two LECs more precisely. After doing so, the energy level $E_0^\gamma$ can be predicted from 
Eq.~\eqref{eq:central} using the PQChPT values of $\{a_0^\gamma,r_0^\gamma\}$. The comparison between the theory prediction and the lattice-extracted value of $E_0^\gamma$ then serves as a concrete accuracy check of the lattice calculation of the more difficult rectangular correlation function $\tilde{C}^R(\tau)$. The last energy level $E_0^\delta$ can be deduced in the same way, but it is just the ordinary $I=0$ energy level that does not depend on any unphysical LECs and can be worked out in the ordinary SU(2) ChPT.

\section{Numerical Analysis}
\label{sec:numerics}

In this section we shall carry out explicitly the procedures outlined in the last paragraph to demonstrate their feasibility. Any lattice working group that has $I=2$ $\pi\pi$ scattering data off the shelf may, however, easily repeat the whole program and even improve upon the precision by performing a more sophisticated error analysis. 

The basis of our analysis below is the lattice data of connected $\pi\pi$ correlation functions $\tilde{C}^{D,C}(\tau)$ kindly provided to us by the \textit{European Twisted Mass} (ETM) Collaboration with $N_f=2+1+1$ and lattice spacing $a\approx 0.086$ fm \cite{Baron:2010bv,Helmes:2015gla}. To study the volume-dependence, we acquire data from two ensembles that share the same bare QCD parameters but with different volumes: A40.32 and A40.24 which correspond to $(L/a)^3\times T/a=32^3\times 64$ and $24^3\times 48$, respectively. The first step is to extract the energy levels $E_0^{\alpha,\beta}$ from the connected $\pi\pi$ correlation functions. In order to minimize pollutions from thermal states, we adopt the technique introduced in Ref.~\cite{Feng:2009ij} and define the function
\begin{equation}
R_i(\tau+a/2)\equiv\frac{\tilde{C}^i(\tau)-\tilde{C}^i(\tau+a)}{C_\pi^2(\tau)-C_\pi^2(\tau+a)}, \label{eq:R_function}
\end{equation}
($i=\alpha,\beta$), where $C_\pi(\tau)$ is the single pion two-point correlation function. We then fit it to the functional form~\cite{Feng:2009ij}
\begin{equation}
R_i(\tau+a/2)=A\left(\cosh(\delta E_0^i \tau')+\sinh(\delta E_0^i\tau')\coth(2M_\pi \tau')\right), 
\end{equation}
where $\tau'=\tau+a/2-T/2$, to extract the energy shift $\delta E_0^i=E_0^i-2M_\pi$. The pion masses at finite volume are taken directly from 
Ref.~\cite{Helmes:2015gla}: $aM_\pi=0.1415(2)$ and $0.1446(3)$ for the ensemble A40.32 and A40.24, respectively. 

\begin{figure}[tb]
	\begin{centering}
		\includegraphics[width=0.49\textwidth]{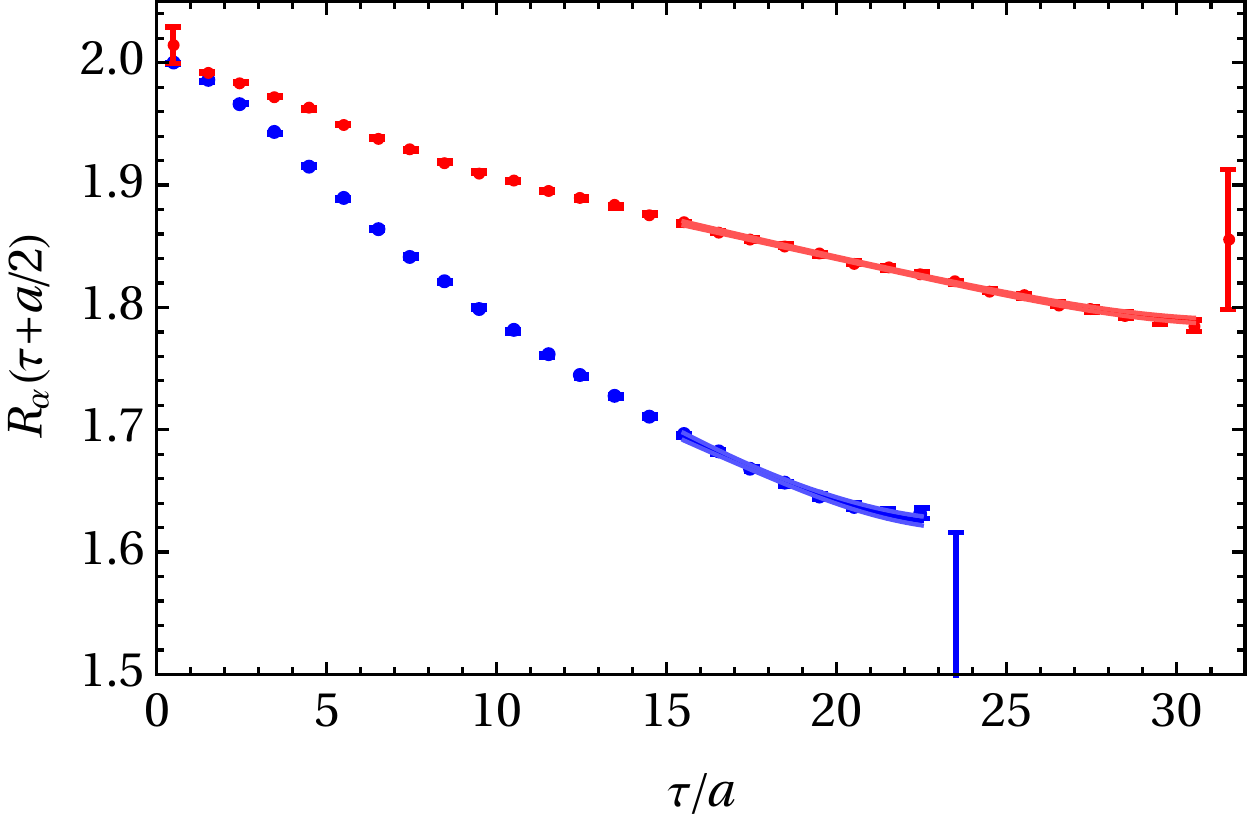}\hfill
		\includegraphics[width=0.49\textwidth]{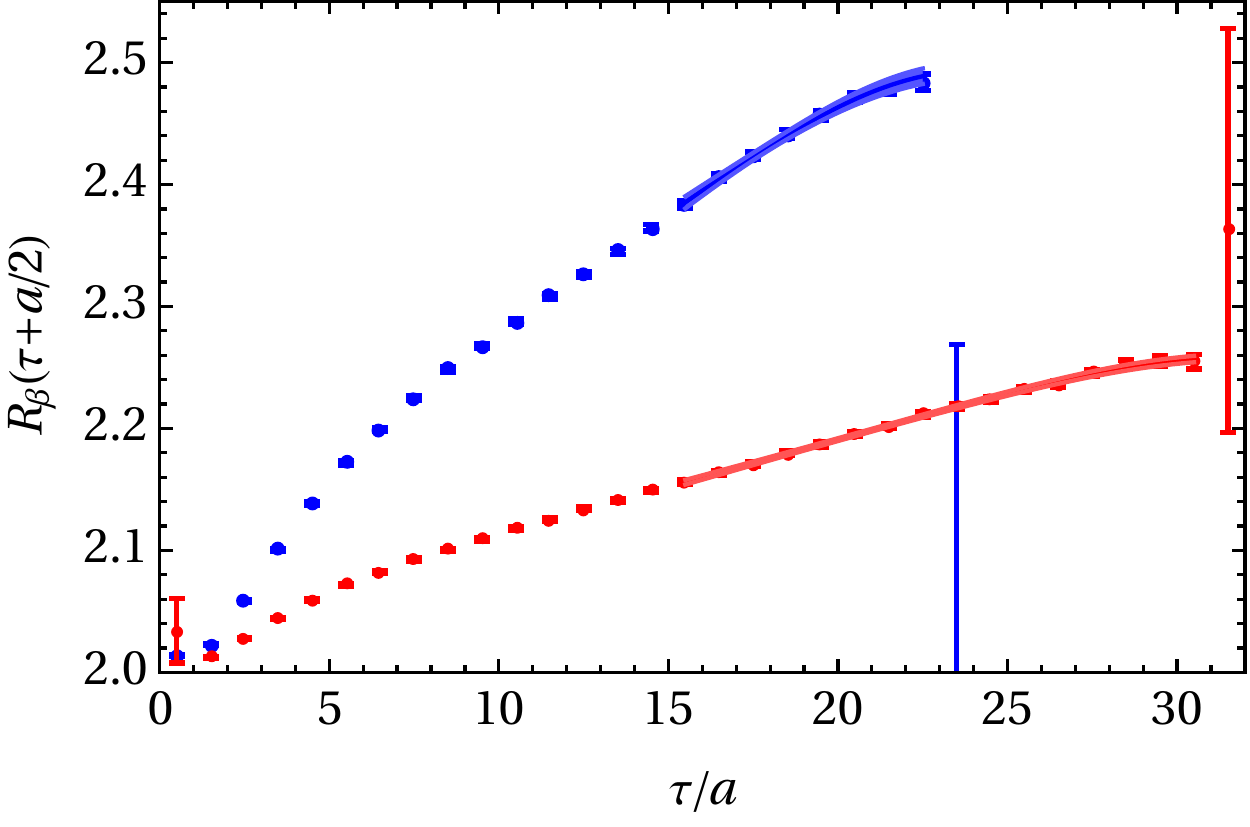}
		\par\end{centering}
	\caption{\label{fig:fit}Fit to the function $R_i(\tau+a/2)$ defined in Eq.~\eqref{eq:R_function} for $i=\alpha$ (left) and $\beta$ (right). Blue and red dots represent the data from the ensemble A40.24 and A40.32 of Ref.~\cite{Helmes:2015gla}, respectively. }
\end{figure}

The error of the fitted energy levels are estimated using the bootstrap method. The distribution of mean values for 1500 bootstrap samples is plotted in Fig.~\ref{fig:fit}, and the error is taken as the standard deviation of the distribution. One extra complication is that the result of fitting depends on the fit range, and in principle one should average over all the results obtained from a large number of different fit ranges 
with a carefully-assigned weight as done in Ref.~\cite{Helmes:2015gla}. In this work, however, we take a shortcut by performing only a single fit for each ensemble over a specially chosen fit range which reproduces best the $I=2$ energy shifts in Ref. ~\cite{Helmes:2015gla}: $16a\leq \tau\leq 31a$ for A40.32 and $16a\leq \tau\leq 23a$ for A40.24. The best-fit curves are displayed in Fig.~\ref{fig:fit} for illustration. The fitted energy levels and their implied threshold parameters are summarized in Table~\ref{tab:fitresult}.
One observes that the scattering lengths are quite well-determined whereas the effective ranges bear a large uncertainty, which is expected because the second term in the effective range expansion, Eq.~\eqref{eq:EFexpand}, has smaller effects on the discrete energy levels, as also observed in Ref.~\cite{Helmes:2015gla}. As a consistency check, we also include in the same table the $I=2$ threshold parameters predicted in the NLO ChPT using Eq.~\eqref{eq:aralpha} (with the values of physical LECs in Table~\ref{tab:LECs}).\footnote{A technical detail: the $M_\pi$ and $F_\pi$ appear in the ChPT expressions refer to the values in the infinite-volume limit; they can be obtained from their finite-volume values divided by the ``finite-volume correction factor" in Table~7 of Ref.~\cite{Helmes:2015gla}. After combining the A40.24 and A40.32 values we obtain $aM_\pi=0.14102(42)$ in lattice units or around 323 MeV, and $M_\pi/F_\pi=2.876(14)$ at $L\rightarrow\infty$.} We observe a slight discrepancy between the NLO ChPT-prediction of the $I=2$ scattering length and the lattice fit at the level of 2.4~sigma,
which could originate from systematic uncertainties not accounted for in the lattice fitting of $\bar{l}_1^\mathrm{phys}$ and $\bar{l}_2^\mathrm{phys}$ \cite{Baron:2010bv}, as they do not enter the pion mass and decay constant at NLO in the infinite-volume ChPT and can only be determined through finite-size effects.

\begin{table}
		\begin{ruledtabular}
		\begin{tabular}{|c |c c c c| }
		 Correlation function& $a\, \delta E_0 $ (A40.32) & $a\, \delta E_0 $ (A40.24) &  { $a_0/a$} & $r_0 /a$ \\ \hline 
		$\tilde{C}^{I=2}$~\cite{Helmes:2015gla} & $\phantom{-}$0.0033(1) & $\phantom{-}$0.0082(3) & $\phantom{-}${ 1.09(6)} & 53(107) \\
		 $\tilde{C}^{I=2}$, SU(2) ChPT & & & 1.300(19) & 83(2)\\
        $\tilde{C}^\alpha$ & $\phantom{-}$0.0034(1) & $\phantom{-}$0.0083(3) & $\phantom{-}${ 1.124(54)} &  41(97) \\
        $\tilde{C}^\beta$ & $-0.0036(1)$ & $-0.0086(3)$ &  $-1.429(77)$ &  140(85) \\
		\end{tabular}
		\end{ruledtabular}
	\caption{The fitted energy shifts, the extracted inverse scattering lengths and effective ranges (with a pion mass in  infinite volume of about 323~MeV) obtained using the connected $\pi\pi$
		correlation functions. Notice that $\delta E_0^\alpha$ is just the $I=2$ $\pi\pi$ energy shift. For comparison, we list the $I=2$ values with the corresponding statistical errors in the original lattice paper~\cite{Helmes:2015gla} in the second row, and include the NLO ChPT predictions of the $I=2$ threshold parameters in the third row. \label{tab:fitresult} }	
		
\end{table}

\begin{table}
	\begin{centering}
		\begin{tabular}{|c|c|c|}
			\hline
			Parameters & Previous & This Work\\ 
			\hline 
		    $F_0$ [MeV] &  85.58(38)& \\
			$\bar{l}_1^\mathrm{phy}$ &  $-0.309(139)$& \\
			$\bar{l}_2^\mathrm{phy}$ & 4.325(10)&\\
			$\bar{l}_3^\mathrm{phy}$ & 3.537(47)&\\
			$\bar{l}_4^\mathrm{phy}$ & 4.735(17)&\\
			$10^3L_{5}^{\mathrm{PQ},r}$ & 0.501(43)& \\
			$10^3L_{8}^{\mathrm{PQ},r}$ &  0.581(22)&\\
			$10^3(3L_{0}^{\mathrm{PQ},r}+L_{3}^{\mathrm{PQ},r})$ &$-0.6(1.4)$ & $-0.70(18)$ \\
			$10^3L_{0}^{\mathrm{PQ},r}$ & 1.0(1.1)& $5.7(1.9)$  \\		
			\hline 
			\hline 
		\end{tabular}
		\par\end{centering}
	\caption{\label{tab:LECs}Parameters in SU$(4|2)$ PQChPT relevant to this work. In the second column, values for the pion decay constant $F_0$ in the chiral limit and the physical LECs 
		$\{\bar{l}_i^\mathrm{phy}\}$ are taken from Ref.~\cite{Baron:2010bv}, whereas 
		$\{L^{\mathrm{PQ,r}}_5,L^{\mathrm{PQ,r}}_8\}$ and $\{3L^{\mathrm{PQ,r}}_0+L_{3}^{\mathrm{PQ},r},L^{\mathrm{PQ,r}}_0\}$ are taken from Ref.~\cite{Boyle:2015exm} through NLO and NNLO fits respectively. In the third column we present our fit of $\{3L^{\mathrm{PQ,r}}_0+L_{3}^{\mathrm{PQ},r},L^{\mathrm{PQ,r}}_0\}$. The renormalization scale of the unphysical LECs is 1 GeV.}	
\end{table}

After evaluating $a_0^\beta$ and $r_0^\beta$, one may proceed to fit the unphysical LECs $\{L_{0,3}^{\mathrm{PQ,r}}\}$ through the following equations:
\begin{eqnarray}
3L_0^\mathrm{PQ,r}+L_3^\mathrm{PQ,r}&=&-\frac{1}{512\pi M_\pi a_0^\beta}-\frac{F_\pi^2}{32M_\pi^2}\left(1+4\mu_\pi\right)+\frac{\bar{l}_1}{384\pi^2}+\frac{\bar{l}_2}{192\pi^2}-\frac{\bar{l}_3}{1024\pi^2}-\frac{\bar{l}_4}{256\pi^2}\nonumber\\
&&+L_5^\mathrm{PQ,r}-L_8^\mathrm{PQ,r}-\frac{3}{1024\pi^2}, \nonumber\\
L_0^\mathrm{PQ,r}&=&\frac{1}{2048\pi}\left(-\frac{3}{M_\pi a_0^\beta}+M_\pi r_0^\beta\right)+ \frac{F_\pi^2 \mu_\pi}{6M_\pi^2}+\frac{\bar{l}_2}{384\pi^2}+\frac{\bar{l}_3}{1024\pi^2}-\frac{1}{2}L_5^\mathrm{PQ,r}\nonumber\\
&&+L_8^\mathrm{PQ,r}+\frac{17}{6144\pi^2}.
\end{eqnarray}
Special attention is paid to the combination $3L_0^\mathrm{PQ,r}+L_3^\mathrm{PQ,r}$ because it is the combination that enters the scattering lengths $\{a_0^\beta,a_0^\gamma\}$, and it is free from the large uncertainty in the effective range $r_0^\beta$. The results are summarized in Table~\ref{tab:LECs}.
One observes that for the quantity $3L_0^\mathrm{PQ,r}+L_0^\mathrm{PQ,r}$, our extraction is in a perfect agreement with the NNLO fit to different quantities by 
Ref.~\cite{Boyle:2015exm}, but with the uncertainty reduced by one order of magnitude. This represents a significant triumph in the PQChPT analysis of $\pi\pi$ scattering. At the same time, a disagreement is found between our extracted $L_0^\mathrm{PQ,r}$ and that obtained in 
Ref.~\cite{Boyle:2015exm} at the level of 1.6 sigma. Here we would like to stress that our analysis is consistently performed at NLO, while the NNLO fits in Ref. \cite{Boyle:2015exm} are known to be more unstable and the results depend quite sensitively on the fitting scheme. We thus believe that our result represents a more realistic estimation of $L_0^\mathrm{PQ,r}$.

\begin{figure}[tb]
	\begin{centering}
		\includegraphics[width=0.49\textwidth]{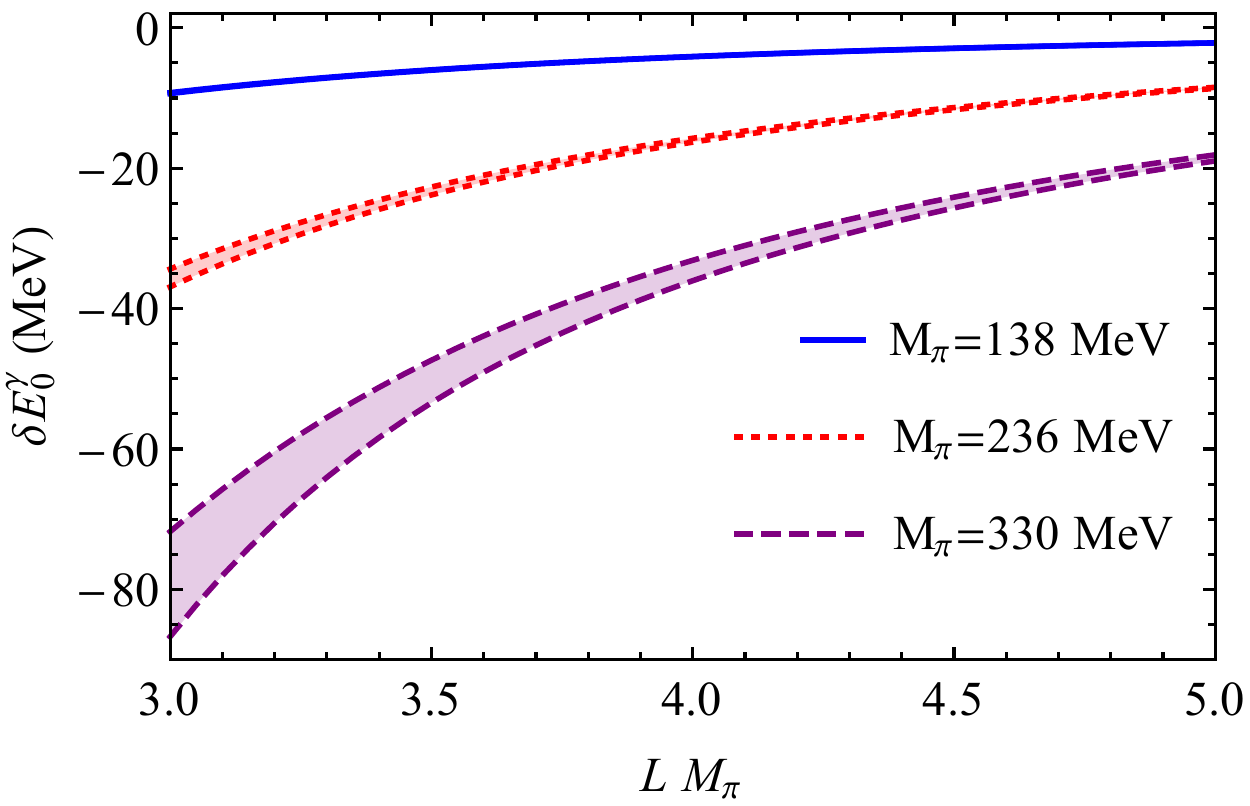}\hfill
		\includegraphics[width=0.49\textwidth]{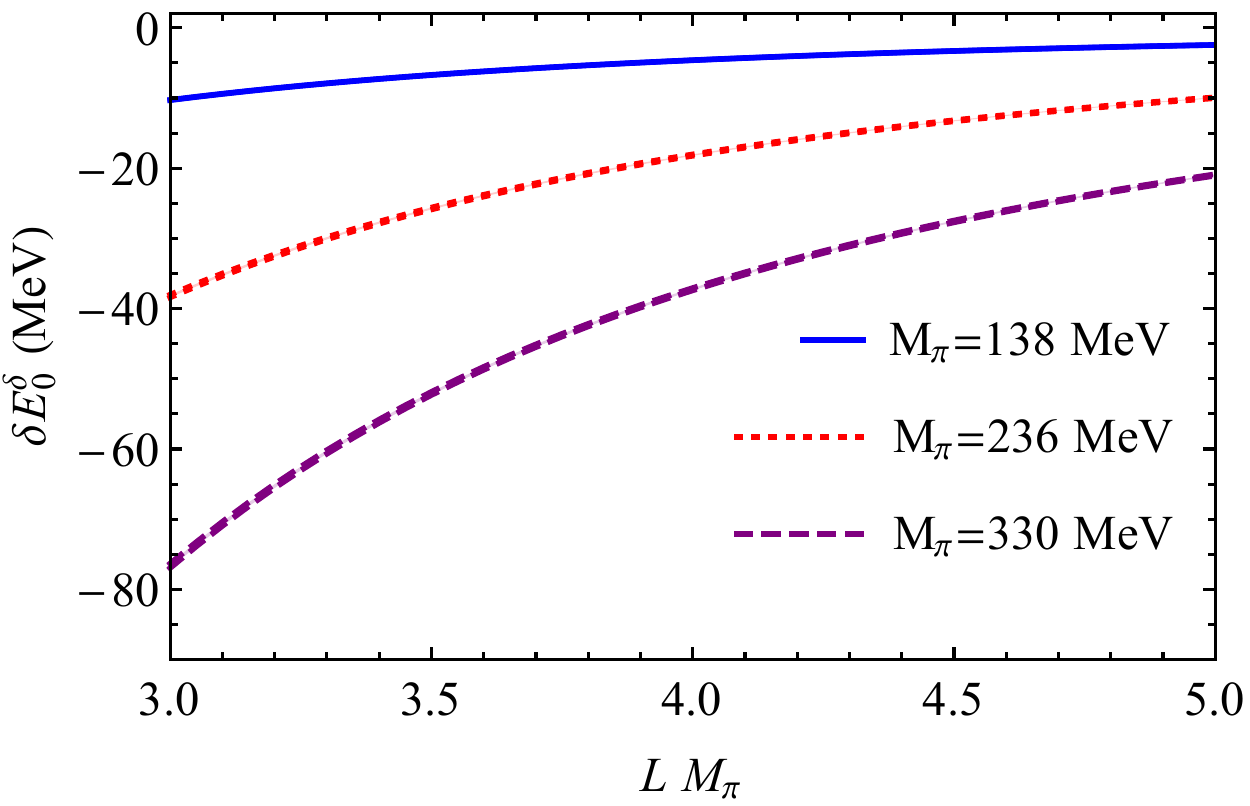}
		\par\end{centering}
	\caption{\label{fig:Egamma}Prediction of the energy shift $\delta E_0^\gamma$ (left) and $\delta E_0^\delta$ (right) as a function of the lattice size $L$ at $M_\pi=$~138 MeV (blue), 236~MeV (red) and 330~MeV (purple). {\color{black} The definitions of $\delta E_0^\gamma$ and $\delta E_0^\delta$ can be found in Eq.~\eqref{eq:Edef}: $\gamma$ corresponds to the channel involving D, C and R types of contractions, and $\delta$ is the isoscalar channel which has all of the D, C, R and V types of contractions. } }
\end{figure}

Finally, we present the most important prediction of this paper, namely the energy shifts $\{\delta E_0^{\gamma,\delta}\}$ as a function of the spatial lattice size $L$ (to remind the readers, the $\delta$-type correlation function is simply the $I=0$ $\pi\pi$ correlation function, while the $\gamma$-type is exactly the same quantity but with vacuum-diagram contributions taken away). To do that, one first calculates the threshold parameters using Eqs.~\eqref{eq:argamma},\eqref{eq:ardelta} and the updated LECs in 
Table~\ref{tab:LECs}.\footnote{The quantity $F_\pi$ in the analytical expressions refers to the physical pion decay constant which is a function of the pion mass. It should be in principle determined directly from lattice, but here we obtain a nice approximated value by using the NLO ChPT expression:  $F_\pi=F_0\left[1+\bar{l}_4\,M_\pi^2/(16\pi^2F_0^2)\right]$.}
Notice that $a_0^\gamma$ depends only on the combination $3L_0^\mathrm{PQ,r}+L_3^\mathrm{PQ,r}$ which is tightly constrained, so its value is determined by satisfactory precision; on the other hand, $r_0^\gamma$ is exposed to the large uncertainty of $L_0^\mathrm{PQ,r}$ especially for the case of a large pion mass which enhances the counterterm effects. Meanwhile, $\{a_0^\delta,r_0^\delta\}$ depend only on physical LECs and are hence tightly constrained. After that, the energy shift is deduced by solving Eq.~\eqref{eq:central} numerically, and the result is plotted in 
Fig.~\ref{fig:Egamma} for $3<LM_\pi<5$ at three pion masses:  $M_\pi=$138, 236 and 330~MeV. 

Here we briefly discuss the qualitative features of the result. First, both the $\gamma$ and $\delta$ channels are attractive so that the energy shifts are negative, and their magnitudes increase with increasing the pion mass and decreasing the lattice size. 
The uncertainty of $\delta E_0^\gamma$ comes primarily from $r_0^\gamma$, whose role is important only when $\delta E_0^\gamma$ is large (in other words, when the effective range expansion in Eq.~\eqref{eq:central} starts breaking down). Therefore, one can see that even at a relatively large $M_\pi$ of 330~MeV, the error bar in $\delta E_0^\gamma$ already starts to diminish at $LM_\pi>4$ and one arrives at a rather controlled prediction of the energy shift. Predictions as in Fig.~\ref{fig:Egamma} are useful because they provide extra consistency checks to the lattice calculations of disconnected contractions in $\pi\pi$ scattering.  As we stress before, our formalism is based on the single-channel L\"{u}scher formula in  partially-quenched QCD which follows from the $S$-matrix unitarity. Thus, its results, bearing higher-order PQChPT corrections, must be obeyed by any lattice setup that satisfies unitarity.
 
\section{Conclusions}
\label{sec:conclusion}

This work serves as an important follow-up to Refs.~\cite{Guo:2013nja,Acharya:2017zje} which represent the initial attempts in the study of generic features of various contraction diagrams in lattice simulations of the $\pi\pi$ scattering using PQQCD. The general motivation is to provide useful insights from the theory side to the noisy ``disconnected diagrams" that usually require special treatments.  A missing ingredient in those works is the discussion of the relation between quantities predictable in PQChPT and the discrete energy levels directly measured on lattice. This work fills up the gap and presents, for the first time, practically-useful theory constraints on the disconnected contributions to two-pion correlation functions that may now be checked by lattice community working in this direction.

Let us briefly summarize our approach. First, each contraction in $\pi\pi$ scattering can be represented by a physical scattering process between two pNGBs in a SU$(4|2)$ PQQCD. The multi-channel scattering matrix can be diagonalized to obtain four effective single-channel scattering amplitudes, two of which involve only connected contractions and the others contain disconnected pieces. With that, we can immediately relate the threshold parameters of each single-channel scattering amplitude with the corresponding discrete energy levels in a finite volume through the usual single-channel L\"uscher formula. The threshold parameters can be expressed in terms of physical and unphysical LECs in the SU$(4|2)$ PQChPT, which can be fitted to discrete energy levels extracted from the connected $\pi\pi$ correlation functions. In particular, we achieve an order-of-magnitude improvement over the result in the existing literature in the determination of the LECs combination $3L_0^{\mathrm{PQ},r}+L_3^{\mathrm{PQ},r}$. Performing lattice computations using more volumes, one can also improve the precision of the worst known $L_0^{\text{PQ},r}$. With these fitted LECs we are able to predict the discrete energy shifts $\delta E_0^{\gamma,\delta}$, which involve disconnected contractions, as functions of the lattice size.

The method can be extended to other sectors, such as the low-energy pion-nucleon scattering, for which lattice calculations are still scarce. Knowledge of the PQChPT LECs extracted from the connected contributions, together with known physical LECs, would be helpful to quantify the disconnected contribution in these more difficult cases, and thus reduce the related systematic uncertainties.

\bigskip

\noindent {\bf Acknowledgements}

We thank Xu Feng and Akaki Rusetsky for many useful discussions, and are particularly indebted to Carsten Urbach, Liuming Liu, Markus Werner and Martin Ueding for providing the lattice data for $\pi\pi$ scattering of the ETM Collaboration and all the patient explanations.
This work is supported in part by  the DFG (Grant No. TRR110)
and the NSFC (Grant No. 11621131001) through the funds provided
to the Sino-German CRC 110 ``Symmetries and the Emergence of
Structure in QCD", by the NSFC under Grant No. 11747601 and No. 11835015, by the CAS under Grant No. QYZDB-SSW-SYS013 and No. XDPB09, by
the CAS Center for Excellence in Particle Physics (CCEPP),  by the Alexander von Humboldt Foundation through the Humboldt Research Fellowship, by the VolkswagenStiftung (Grant No. 93562), and by  the  CAS  President's  International
Fellowship  Initiative  (PIFI)  (Grant  No. 2018DM0034).

\begin{appendix}

\end{appendix}

\end{document}